\documentclass[pdftex,twocolumn,epjc3]{svjour3}          

\RequirePackage[T1]{fontenc}

\smartqed  

\RequirePackage{graphicx}
\RequirePackage{mathptmx}      
\RequirePackage{flushend}
\RequirePackage[numbers,sort&compress]{natbib}
\RequirePackage[colorlinks,citecolor=blue,urlcolor=blue,linkcolor=blue]{hyperref}

\journalname{Eur. Phys. J. C}
\usepackage{epsfig,amsmath,amssymb}
\usepackage{color}
\usepackage{float}
\usepackage{subfigure}
\usepackage[utf8]{inputenc}

\begin{document}

\title{A novel and economical explanation for SM fermion masses and mixings}


\author{A. E. C\'arcamo Hern\'andez \thanksref{e1,addr1}}
\thankstext{e1}{antonio.carcamo@usm.cl}
 \institute{Universidad T\'{e}cnica Federico Santa Mar\'{\i}a, Casilla 110-V, Valpara\'{\i}so, Chile \label{addr1}}
\date{Received: 16th of June 2016/Accepted: 2th of September 2016}

\maketitle

\begin{abstract}
I propose the first multiscalar singlet extension of the Standard Model
(SM), that generates tree level top quark and exotic fermion masses as well
as one and three loop level masses for charged fermions lighter than the top
quark and for light active neutrinos, respectively, without invoking
electrically charged scalar fields. That model, which is based on the $S_{3}\times Z_{8}$ discrete symmetry, successfully explains the observed SM fermion mass and mixing pattern. The charged exotic fermions induce one loop level masses for charged fermions lighter than the top quark. The $Z_{8}$ charged scalar singlet $\chi $ generates the observed charged fermion mass and quark mixing pattern. 
\end{abstract}


\thankstext{e1}{antonio.carcamo@usm.cl} 
\institute{Universidad T\'{e}cnica
Federico Santa Mar\'{\i}a and Centro Cient\'{\i}fico-Tecnol\'{o}gico de
Valpara\'{\i}so, Casilla 110-V, Valpara\'{\i}so, Chile \label{addr1}}


\section{Introduction}

Despite its great consistency with the
experimental data, the Standard Model (SM) is unable
to explain several issues such as, for example, the number
of fermion generations, the observed pattern of fermion
masses and mixings, etc. In this letter I propose the first multiscalar singlet extension of the SM, that generates tree level top quark and exotic fermion masses as well as one and three loop level masses for charged fermions lighter than the top quark and for light active neutrinos, respectively, without invoking electrically
charged scalar fields. That multiscalar singlet extension is consistent with
the SM fermion mass and mixing pattern. 

\section{The Model}

The model has the SM gauge symmetry, which is supplemented by the $%
S_{3}\times Z_{8}$ discrete group. It is noteworthy that among the discrete
symmetries, I introduced the symmetry group $S_{3}$ since it is the smallest
non-Abelian group that has been considerably studied in the literature. The $%
S_{3}$ symmetry is assumed to be preserved whereas the $Z_{8}$ discrete
group is broken at the scale $v_{\chi }$. The breaking of the $Z_{8}$
symmetry gives rise to the observed charged fermion mass and quark mixing
pattern. The scalar sector of the SM is extended by introducing three EW
scalar singlets, i.e., $\eta _{1}$, $\eta _{2}$ and $\chi $, assumed to be
charged under the $Z_{8}$ symmetry. Out of these three SM scalar singlets,
two scalar fields ($\eta _{1}$, $\eta _{2}$) are grouped in a $S_{3}$
doublet, namely $\eta $, whereas the remaining one ($\chi $) is assigned to
be a trivial $S_{3}$ singlet. The SM Higgs doublet $\phi $ is assigned to be
a trivial $S_{3}$ singlet, neutral under the $Z_{8}$ discrete symmetry.
Since the $S_{3}$ symmetry is preserved, the $S_{3}$ scalar doublet $\eta
=\left( \eta _{1},\eta _{2}\right) $ does not acquire a vacuum expectation
value. The remaining scalar fields, i.e, $\phi $ and $\chi $, which are
assigned as trivial $S_{3}$ singlets, acquire nonvanishing vacuum
expectation values, as it should be. Regarding the SM fermion sector, I
assign the left handed fermionic fields, the right handed top quark field as
trivial $S_{3}$ singlets and the remaining right handed SM fermionic fields
as $S_{3}$ nontrivial singlets, implying that the top quark is the only SM
charged fermion that acquires a tree level mass. The remaining SM charged
fermions get their masses from a one loop radiative seesaw mechanism and the
hierarchy among their masses will arise from the different $Z_{8}$ charge
assignments of the fermionic fields. As it will be shown in the following,
light active neutrinos masses will arise from a three loop radiative seesaw
mechanism. In order that all fermions lighter than the top quark acquire non
vanishing masses, the fermion sector of the Standard Model is extended by
including: two heavy right handed Majorana neutrinos $\nu _{1R}$, $\nu _{2R}$%
, six SM gauge singlet charged leptons $E_{\kappa L}$ and $E_{\kappa R}$ ($%
\kappa =1,\cdots ,6$), ten heavy $SU(2)_{L}$ singlet exotic quarks $%
B_{\kappa L}$, $B_{\kappa R}$ ($\kappa =1,\cdots ,6$), $T_{\lambda L} $, $%
T_{\lambda R}$ ($\lambda =1,\cdots ,4$). The heavy exotic $T_{\lambda }$ ($%
\lambda =1,\cdots ,4$)\ and $B_{\kappa }$ ($\kappa =1,\cdots ,6$)\ quarks 
should have electric charges equal to $\frac{2}{3}$ and $-\frac{1}{3}$,
respectively, in order to implement a one loop radiative seesaw mechanism
that generates masses for quarks, lighter than the top quark. To build the
Yukawa terms invariant under the $S_{3}$ discrete group, I assign the two
heavy right handed Majorana neutrinos as non trivial $S_{3}$ singlets,
whereas the non SM charged fermions are grouped into the $S_{3}$ doublets $%
T_{L,R}^{\left( r\right) }$ ($r=1,2$), $B_{L,R}^{\left( k\right) }$ and $%
E_{L,R}^{\left( k\right) }$ ($k=1,2,3$). I do not unify the two heavy right
handed Majorana neutrinos in a $S_{3}$ doublet since that assignment will
result in two massless active neutrinos, which is in clear contradiction
with the neutrino oscillation experimental data. The non SM fermionic fields
described above, together with the $S_{3}$ doublet $\eta =\left( \eta
_{1},\eta _{2}\right) $ will induce one and three loop radiative seesaw
mechanisms to generate the masses for fermions lighter than the top quark
and for the light active neutrinos, respectively. The aforementioned non SM
fermion content is the minimal required to generate the masses for fermions
lighter than the top quark and for the light active neutrinos. I further
assume that the fermionic fields transform under the $Z_{8}$ symmetry, as
follows:\vspace{-0.1cm} 
\begin{eqnarray}
{\small q_{jL}} &\rightarrow &e^{-\frac{\pi i\left( 3-j\right) }{2}}{\small %
q_{jL}},\hspace{0.25cm}u{\small _{jR}}\rightarrow e^{\frac{\pi i\left(
3-j\right) }{2}}u{\small _{jR}},\hspace{0.25cm}d{\small _{jR}}\rightarrow e^{%
\frac{\pi i\left( 3-j\right) }{2}}d{\small _{jR}},  \notag \\
l{\small _{jL}} &\rightarrow &e^{-\frac{\pi i\left( 3-j\right) }{2}}l{\small %
_{jL}},\hspace{0.5cm}l{\small _{jR}}\rightarrow e^{\frac{\pi i\left(
3-j\right) }{2}}l{\small _{jR}},\hspace{0.5cm}j=1,2,3,  \notag \\
T_{L}^{\left( r\right) } &\rightarrow &e^{-\frac{\pi i}{4}}T_{L}^{\left(
r\right) },\hspace{0.5cm}T_{R}^{\left( r\right) }\rightarrow e^{\frac{\pi i}{%
4}}T_{R}^{\left( r\right) },\hspace{0.5cm}r=1,2,  \notag \\
B_{L}^{\left( k\right) } &\rightarrow &e^{-\frac{\pi i}{4}}B_{L}^{\left(
k\right) },\hspace{0.5cm}B_{R}^{\left( k\right) }\rightarrow e^{\frac{\pi i}{%
4}}B_{R}^{\left( k\right) },\hspace{0.5cm}k=1,2,3,  \notag \\
E_{L}^{\left( k\right) } &\rightarrow &e^{-\frac{\pi i}{4}}E_{L}^{\left(
k\right) },\hspace{0.5cm}E_{R}^{\left( k\right) }\rightarrow e^{\frac{\pi i}{%
4}}E_{R}^{\left( k\right) },\hspace{0.5cm}k=1,2,3,  \notag \\
\nu _{sR} &\rightarrow &e^{-\frac{\pi i}{4}}\nu _{sR},\hspace{0.5cm}s=1,2.
\label{fermionassignments}
\end{eqnarray}
The EW scalar singlets $\eta =\left( \eta _{1},\eta _{2}\right) $ and $\chi $
are charged under the $Z_{8}$ symmetry as 
$\eta \rightarrow e^{-\frac{\pi i}{4}}\eta, \chi \rightarrow e^{-\frac{\pi i%
}{2}}\chi$. 
With the above particle content, the following quark and charged lepton and
neutrino Yukawa terms invariant under the symmetries of the model arise:%
\vspace{-0.2cm} 
\begin{eqnarray}
-\mathcal{L}_{\text{Y}}^{\left( u\right) }
&=&\sum_{j=1}^{3}\sum_{r=1}^{2}y_{jr}^{\left( u\right) }\overline{q}_{jL}%
\widetilde{\phi }\left( T_{R}^{\left( r\right) }\eta \right) _{\mathbf{1}}%
\frac{\chi ^{3-j}}{\Lambda ^{4-j}} \\
&+&\sum_{r=1}^{2}\sum_{s=1}^{2}x_{rs}^{\left( u\right) }\left( \overline{T}%
_{L}^{\left( r\right) }\eta\right) _{\mathbf{1}^{\prime }}u_{sR}\frac{\chi
^{3-k}}{\Lambda ^{3-k}}  \notag \\
&+&\sum_{j=1}^{3}y_{j3}^{\left( u\right) }\overline{q}_{jL}\widetilde{\phi }%
u_{3R}\frac{\chi ^{3-j}}{\Lambda ^{3-j}}+\sum_{r=1}^{2}y_{r}^{\left(
T\right) }\left( \overline{T}_{L}^{\left( r\right) }T_{R}^{\left( r\right)
}\right) _{\mathbf{1}}\chi+h.c  \notag
\end{eqnarray}%
\vspace{-0.4cm} 
\begin{eqnarray}
-\mathcal{L}_{\text{Y}}^{\left( d\right) }
&=&\sum_{j=1}^{3}\sum_{k=1}^{3}y_{jk}^{\left( d\right) }\overline{q}%
_{jL}\phi \left( B_{R}^{\left( k\right) }\eta \right) _{\mathbf{1}}\frac{%
\chi ^{3-j}}{\Lambda ^{4-j}}  \notag \\
&+&\sum_{j=1}^{3}\sum_{k=1}^{3}x_{jk}^{\left( d\right) }\left( \overline{B}%
_{L}^{\left( j\right) }\eta\right) _{\mathbf{1}^{\prime }}d_{kR}\frac{\chi
^{3-k}}{\Lambda ^{3-k}}  \notag \\
&+&\sum_{k=1}^{3}y_{k}^{\left( B\right) }\left( \overline{B}_{L}^{\left(
k\right) }B_{R}^{\left( k\right) }\right) _{\mathbf{1}}\chi +h.c.
\label{Lyq}
\end{eqnarray}%
\vspace{-0.4cm} 
\begin{eqnarray}
-\mathcal{L}_{\text{Y}}^{\left( l\right) }
&=&\sum_{j=1}^{3}\sum_{k=1}^{3}y_{jk}^{\left( l\right) }\overline{l}%
_{jL}\phi \left( E_{R}^{\left( k\right) }\eta \right) _{\mathbf{1}}\frac{%
\chi ^{3-j}}{\Lambda ^{4-j}}  \notag \\
&+&\sum_{j=1}^{3}\sum_{k=1}^{3}x_{jk}^{\left( l\right) }\left( \overline{E}%
_{L}^{\left( j\right) }\eta\right) _{\mathbf{1}^{\prime }}l_{kR}\frac{\chi
^{3-k}}{\Lambda ^{3-k}}  \notag \\
&+&\sum_{k=1}^{3}y_{k}^{\left( E\right) }\left( \overline{E}_{L}^{\left(
k\right) }E_{R}^{\left( k\right) }\right) _{\mathbf{1}}\chi+h.c  \label{Lyl2}
\end{eqnarray}%
\vspace{-0.2cm} 
\begin{eqnarray}
-\mathcal{L}_{\text{Y}}^{\left( \nu \right)
}&=&\sum_{j=1}^{3}\sum_{s=1}^{2}y_{js}^{\left( \nu \right) }\overline{l}_{jL}%
\widetilde{\phi }\nu _{sR}\frac{\left[ \eta ^{\ast }\left( \eta \eta ^{\ast
}\right) _{\mathbf{2}}\right] _{\mathbf{1}^{\prime }}\chi ^{3-j}}{\Lambda
^{6-j}}  \notag \\
&+&\sum_{s=1}^{2}y_{s}\overline{\nu }_{sR}\nu _{sR}^{C}\chi +h.c.
\label{Lynu}
\end{eqnarray}
Where, for the sake of simplificity, I have neglected the mixing terms
between the different $S_{3}$ fermionic doublets as well as the mixings
between the right handed Majorana neutrinos. I have assumed that the non SM
fermions are physical fields. After the spontaneous breaking of the
electroweak and the $Z_{8}$ discrete symmetry and considering that the $%
S_{3} $ symmetry is preserved, the Yukawa interactions given above will
generate tree level masses for the top quark and for the non SM fermions and
one loop level masses for the remaining SM charged fermions. To generate the
one loop level masses for the charged fermions lighter than the top quark,
the $S_{3}$ symmetry has to be softly broken by adding a $\mu
_{12}^{2}\eta_{1}\eta _{2} $ term in the scalar potential for the $S_{3}$
scalar doublet $\eta =\left( \eta _{1},\eta _{2}\right) $. Since the $S_{3}$
scalar doublet $\eta =\left( \eta _{1},\eta _{2}\right) $ has a vanishing
vacuum expectation value, light active neutrinos do not acquire tree level
masses, they get masses via a three loop radiative seesaw mechanism (as
follows from the $S_{3}$ invariance of the neutrino Yukawa interactions)
that involves the two heavy Majorana neutrinos $\nu _{1R}$, $\nu _{2R}$ as
well as the real and imaginary parts of the $\eta _{1}$, $\eta _{2}$ scalar
fields running in the loops. It is remarkable that this three loop level
radiative seesaw mechanism for light active neutrino masses does not require
charged scalar fields as in the other three loop level mechanisms discussed
in the literature \cite{Ma:2015xla}.\newline
Since the hierarchy of charged fermion masses and quark mixing angles arises
from the breaking of the $Z_{8}$ discrete group, and in order to relate the
quark masses with the quark mixing parameters, the vacuum expectation value
(VEV) of the SM scalar singlet $\chi $ is set as follows: $v_{\chi }=\lambda
\Lambda$, where $\lambda =0.225$ is one of the Wolfenstein parameters and $%
\Lambda $ corresponds to the model cutoff.

\section{Fermion masses and mixings}

From the Yukawa terms, it follows that the quark, charged lepton and light
active neutrino mass matrices have the form:\vspace{-0.15cm} 
\begin{equation}
M_{U}=\left( 
\begin{array}{ccc}
\varepsilon _{11}^{\left( u\right) }\lambda ^{3} & \varepsilon _{12}^{\left(
u\right) }\lambda ^{2} & y_{13}^{\left( u\right) }\lambda ^{2} \\ 
\varepsilon _{21}^{\left( u\right) }\lambda ^{2} & \varepsilon _{22}^{\left(
u\right) }\lambda  & y_{23}^{\left( u\right) }\lambda  \\ 
\varepsilon _{31}^{\left( u\right) }\lambda  & \varepsilon _{32}^{\left(
u\right) } & y_{33}^{\left( u\right) }%
\end{array}%
\right) \frac{v}{\sqrt{2}},
\end{equation}%
\begin{eqnarray}
M_{D,l} &=&\left( 
\begin{array}{ccc}
\varepsilon _{11}^{\left( d,l\right) }\lambda ^{4} & \varepsilon
_{12}^{\left( d,l\right) }\lambda ^{3} & \varepsilon _{13}^{\left(
d,l\right) }\lambda ^{2} \\ 
\varepsilon _{21}^{\left( d,l\right) }\lambda ^{3} & \varepsilon
_{22}^{\left( d,l\right) }\lambda ^{2} & \varepsilon _{23}^{\left(
d,l\right) }\lambda  \\ 
\varepsilon _{31}^{\left( d,l\right) }\lambda ^{2} & \varepsilon
_{32}^{\left( d,l\right) }\lambda  & \varepsilon _{33}^{\left( d,l\right) }%
\end{array}%
\right) \frac{v}{\sqrt{2}},  \label{Md} \\
M_{\nu } &=&\left( 
\begin{array}{ccc}
W_{1}^{2} & W_{1}W_{2}\cos \varphi  & W_{1}W_{3}\cos \left( \varphi -\varrho
\right)  \\ 
W_{1}W_{2}\cos \varphi  & W_{2}^{2} & W_{2}W_{3}\cos \varrho  \\ 
W_{1}W_{3}\cos \left( \varphi -\varrho \right)  & W_{2}W_{3}\cos \varrho  & 
W_{3}^{2}%
\end{array}%
\right) ,  \notag
\end{eqnarray}
where: 
\begin{eqnarray}
\overrightarrow{W_{j}} &=&\left( \frac{A_{j1}\sqrt{y_{1}v_{\chi }f_{1}^{\left(
\nu \right) }}}{64\pi ^{3}\Lambda },\frac{A_{j2}\sqrt{y_{2}v_{\chi}f_{2}^{\left(
\nu \right) }}}{64\pi ^{3}\Lambda }\right) ,\hspace{0.5cm}j=1,2,3\notag \\
\cos \varphi  &=&\frac{\overrightarrow{W_{1}}\cdot \overrightarrow{W_{2}}}{%
\left\vert \overrightarrow{W_{1}}\right\vert \left\vert \overrightarrow{W_{2}%
}\right\vert },\hspace{0.25cm}\cos \left( \varphi -\varrho \right) =\frac{%
\overrightarrow{W}_{1}\cdot \overrightarrow{W_{3}}}{\left\vert 
\overrightarrow{W}_{1}\right\vert \left\vert \overrightarrow{W_{3}}%
\right\vert },\hspace{0.25cm}W_{j}=\left\vert \overrightarrow{W_{j}}%
\right\vert,\notag\\
\cos \varrho  &=&\frac{\overrightarrow{W_{2}}\cdot \overrightarrow{W_{3}}}{%
\left\vert \overrightarrow{W_{2}}\right\vert \left\vert \overrightarrow{W_{3}%
}\right\vert },\hspace{0.5cm}A_{js}=\lambda ^{3-j}y_{js}^{\left( \nu \right) }\frac{v}{\sqrt{2}}\hspace{0.5cm}s=1,2,
\end{eqnarray}
where $\varepsilon _{jk}^{\left( f\right) }$ ($j,k=1,2,3$) with $f=u,d,l$
and $f_{s}^{\left( \nu \right) }$ ($s=1,2$) are dimensionless parameters
generated at one and three loop levels, respectively. Furthermore, the
entries of the charged lepton and down type quark mass matrices exhibit the
same scalings in terms of the Wolfenstein parameter $\lambda $, as seen from
Eqs. (\ref{Md}). 
Since the dimensionless parameters $\varepsilon _{jk}^{\left( f\right) }$ ($%
j,k=1,2,3$) with $f=u,d,l$, are generated at one loop level, I set $%
\varepsilon _{jk}^{\left( f\right) }=a_{jk}^{\left( f\right) }\lambda ^{3}$,
where $a_{jk}^{\left( f\right) }$ are $\mathcal{O}(1)$ parameters. Since the
charged fermion mass and quark mixing angles is caused by the breaking of
the $Z_{8}$ discrete group and in order to simplify the analysis, I adopt a
benchmark where I set:
\begin{eqnarray}
a_{12}^{\left( u\right) } &=&a_{21}^{\left( u\right) },\hspace{0.7cm}%
a_{31}^{\left( u\right) }=y_{13}^{\left( u\right) },\hspace{0.7cm}%
a_{32}^{\left( u\right) }=y_{23}^{\left( u\right) }\notag \\
a_{12}^{\left( d\right) } &=&\left\vert a_{12}^{\left( d\right) }\right\vert
e^{-i\tau _{1}},\hspace{0.75cm}a_{21}^{\left( d\right) }=\left\vert
a_{12}^{\left( d\right) }\right\vert e^{i\tau _{1}}, \\
a_{13}^{\left( d\right) } &=&\left\vert a_{13}^{\left( d\right) }\right\vert
e^{-i\tau _{2}},\hspace{0.75cm}a_{31}^{\left( d\right) }=\left\vert
a_{13}^{\left( d\right) }\right\vert e^{i\tau _{2}},\hspace{0.75cm}%
a_{23}^{\left( d\right) }=a_{32}^{\left( d\right) }\notag.
\end{eqnarray}
Furthermore I set $a_{33}^{\left( u\right) }=1$, as suggested by naturalness
arguments. Then, I proceed to fit the effective parameters $a_{11}^{\left(
u\right) }$, $a_{22}^{\left( u\right) }$, $a_{12}^{\left( u\right) }$, $%
a_{13}^{\left( u\right) }$, $a_{23}^{\left( u\right) }$, $a_{11}^{\left(
d\right) }$, $a_{22}^{\left( d\right) }$, $a_{33}^{\left( d\right) }$, $%
\left\vert a_{12}^{\left( d\right) }\right\vert $, $\left\vert
a_{13}^{\left( d\right) }\right\vert $, $a_{23}^{\left( d\right) }$ and the
phases $\tau _{1}$, $\tau _{2}$, to reproduce the experimental values of the quark masses, the three quark mixing angles and the CP violating phase $\delta $. Their obtained values 
in Table \ref{Tab} correspond to the best fit values: 
\begin{eqnarray}
a_{11}^{\left( u\right) } &\simeq &0.58,\hspace{1cm}a_{22}^{\left( u\right)
}\simeq 2.19,\hspace{1cm}a_{12}^{\left( u\right) }\simeq 0.67,\notag \\
a_{13}^{\left( u\right) } &\simeq &0.80,\hspace{1cm}a_{23}^{\left( u\right)
}\simeq 0.83,\hspace{1cm}a_{11}^{\left( d\right) }\simeq 1.96,\notag \\
a_{12}^{\left( d\right) } &\simeq &0.53,\hspace{1cm}a_{13}^{\left( d\right)
}\simeq 1.07,\hspace{1cm}a_{22}^{\left( d\right) }\simeq 1.93, \\
a_{23}^{\left( d\right) } &\simeq &1.36,\hspace{0.5cm}a_{33}^{\left(
d\right) }\simeq 1.35,\hspace{0.5cm}\tau _{1}\simeq 9.56^{\circ },\hspace{%
0.5cm}\tau _{2}\simeq 4.64^{\circ }\notag.
\end{eqnarray}%
\begin{table}[tbh]
\begin{center}
\begin{tabular}{c|l|l}
\hline\hline
Observable & Model value & Experimental value \\ \hline
$m_{u}(MeV)$ & \quad $1.44$ & \quad $1.45_{-0.45}^{+0.56}$ \\ \hline
$m_{c}(MeV)$ & \quad $656$ & \quad $635\pm 86$ \\ \hline
$m_{t}(GeV)$ & \quad $177.1$ & \quad $172.1\pm 0.6\pm 0.9$ \\ \hline
$m_{d}(MeV)$ & \quad $2.9$ & \quad $2.9_{-0.4}^{+0.5}$ \\ \hline
$m_{s}(MeV)$ & \quad $57.7$ & \quad $57.7_{-15.7}^{+16.8}$ \\ \hline
$m_{b}(GeV)$ & \quad $2.82$ & \quad $2.82_{-0.04}^{+0.09}$ \\ \hline
$\sin \theta _{12}$ & \quad $0.225$ & \quad $0.225$ \\ \hline
$\sin \theta _{23}$ & \quad $0.0412$ & \quad $0.0412$ \\ \hline
$\sin \theta _{13}$ & \quad $0.00351$ & \quad $0.00351$ \\ \hline
$\delta $ & \quad $64^{\circ }$ & \quad $68^{\circ }$ \\ \hline\hline
\end{tabular}%
\end{center}
\caption{Model and experimental values of the quark masses and CKM
parameters.}
\label{Tab}
\end{table}

The obtained quark masses, quark mixing angles and CP violating phase are consistent with the experimental data. 
The obtained and experimental values for the physical observables of the quark
sector are reported in Table \ref{Tab}. I use the experimental values of
the quark masses at the $M_{Z}$ scale, from Ref. \cite{Bora:2012tx}, whereas the experimental values of the CKM parameters are taken from Ref. \cite{Agashe:2014kda}.

In what follows I will explain the reason for choosing the $Z_{8}$ discrete
symmetry. It is noteworthy that the $Z_{4}$ discrete group is the smallest
cyclic symmetry that allows to get the $\lambda ^{2}$ suppression in the 13
entry of the up type quark mass matrix from a $\frac{\chi ^{2}}{\Lambda ^{2}}
$ insertion on the $\overline{q}_{1L}\widetilde{\phi }u_{3R}$ operator.
However, since the masses of the non SM fermions arise from their
renormalizable Yukawa interactions with the SM scalar singlet $\chi $,
charged under the discrete cyclic group, the invariance of these Yukawa
interactions under the cyclic symmetry requires to consider the $Z_{8}$
instead of the $Z_{4}$ discrete symmetry.

In the concerning to the charged lepton sector, I adopt a benchmark where I
set $a_{jk}^{\left(l\right) }=a_{k}^{\left( l\right) }\delta _{jk}$. This
benchmark scenario that leads to a diagonal charged lepton mass matrix is
justified by the requirement of forbidding unobserved lepton flavor
violating processes such as $\mu \rightarrow e\gamma $. This assumption is also made in the framework of an extended inert two Higgs doublet model addressed to explain the charged lepton mass hierarchy \cite{Okada:2013iba}. Within the benchmark previously described, the charged lepton masses take the form:
\begin{equation*}
m_{e}=a_{1}^{\left( l\right) }\lambda ^{7}\frac{v}{\sqrt{2}},\hspace{0.5cm}%
m_{\mu }=a_{2}^{\left( l\right) }\lambda ^{5}\frac{v}{\sqrt{2}},\hspace{0.5cm%
}m_{\tau }=a_{3}^{\left( l\right) }\lambda ^{3}\frac{v}{\sqrt{2}}.
\end{equation*}
Regarding the neutrino sector, from the light active neutrino mass texture $M_{\nu}$, it follows that the light active neutrino spectrum includes one massless neutrino and two massive neutrinos for both normal and inverted neutrino mass hierarchies. It is noteworthy that two heavy Majorana neutrinos are needed in this model to
generate the two nonvanishing neutrino mass squared splittings measured in
neutrino oscillation experiments. Furthermore, due to the preserved $S_{3}$
symmetry, the model allows for stable dark matter candidates, which could be
either the right handed Majorana neutrinos $\nu _{sR}$ or the lightest of
the scalar fields $Re(\eta _{s})$ and $Im(\eta _{s})$ ($s=1,2$). 
 In order to show that the light active neutrino texture $M_{\nu}$ can fit the experimental data, I set $\varphi =2\varrho $ only for the case of normal hierarchy (NH). 
Varying the lepton sector model parameters $a_{1}^{\left(
l\right) }$, $a_{2}^{\left( l\right) }$, $a_{3}^{\left( l\right) }$, $%
\varrho $, $W_{1}$, $W_{2}$ and $W_{3}$ (as well as $\varphi $ for IH only),
I fitted the charged lepton masses, the neutrino mass squared splitings $%
\Delta m_{21}^{2}$, $\Delta m_{31}^{2}$ 
and the leptonic mixing parameters $\sin^{2}\theta _{12}$, $\sin ^{2}\theta _{13}$ and $\sin ^{2}\theta _{23}$\ to their experimental values for NH and IH. The results shown in Table \ref{Observables0} correspond to the following best-fit values: 
\begin{eqnarray}
\varrho  &\simeq &38.73^{\circ },\hspace{0.5cm}W_{1}\simeq -0.063eV^{\frac{1%
}{2}},\hspace{0.5cm}W_{2}\simeq 0.18eV^{\frac{1}{2}},  \notag \\
W_{3} &\simeq &0.15eV^{\frac{1}{2}},\hspace{0.5cm}\mbox{for NH}  \notag \\
a_{1}^{\left( l\right) } &\simeq &0.1,\hspace{0.5cm}a_{2}^{\left( l\right)
}\simeq 1.02,\hspace{0.5cm}a_{3}^{\left( l\right) }\simeq 0.88,\ \ \ \ \ \ \
\   \label{ParameterfitNH}
\end{eqnarray}\vspace{-0.6cm}
\begin{eqnarray}
\varrho  &\simeq &162.26^{\circ },\hspace{0.5cm}\varphi \simeq 79.44^{\circ
},\hspace{0.5cm}W_{1}\simeq 0.22eV^{\frac{1}{2}},  \notag \\
W_{2} &\simeq &0.15eV^{\frac{1}{2}},\hspace{0.5cm}W_{3}\simeq 0.17eV^{\frac{1%
}{2}},\hspace{0.5cm}\mbox{for IH}  \notag \\
a_{1}^{\left( l\right) } &\simeq &0.1,\hspace{0.5cm}a_{2}^{\left( l\right)
}\simeq 1.02,\hspace{0.5cm}a_{3}^{\left( l\right) }\simeq 0.88,
\label{ParameterfitIH}
\end{eqnarray}\vspace{-0.5cm}
\begin{table}[tbh]
\begin{center}
\begin{tabular}{c|l|l}
\hline\hline
Observable & Model value & Experimental value \\ \hline
$m_{e}(MeV)$ & \quad $0.487$ & \quad $0.487$ \\ \hline
$m_{\mu }(MeV)$ & \quad $102.8$ & \quad $102.8\pm 0.0003$ \\ \hline
$m_{\tau }(GeV)$ & \quad $1.75$ & \quad $1.75\pm 0.0003$ \\ \hline
$\Delta m_{21}^{2}$($10^{-5}$eV$^{2}$) (NH) & \quad $7.22$ & \quad $%
7.60_{-0.18}^{+0.19}$ \\ \hline
$\Delta m_{31}^{2}$($10^{-3}$eV$^{2}$) (NH) & \quad $2.50$ & \quad $%
2.48_{-0.07}^{+0.05}$ \\ \hline
$\sin ^{2}\theta _{12}$ (NH) & \quad $0.334$ & \quad $0.323\pm 0.016$ \\ 
\hline
$\sin ^{2}\theta _{23}$ (NH) & \quad $0.567$ & \quad $%
0.567_{-0.128}^{+0.032} $ \\ \hline
$\sin ^{2}\theta _{13}$ (NH) & \quad $0.0228$ & \quad $0.0234\pm 0.0020$ \\ 
\hline
$\Delta m_{21}^{2}$($10^{-5}$eV$^{2}$) (IH) & \quad $7.60$ & \quad $%
7.60_{-0.18}^{+0.19}$ \\ \hline
$\Delta m_{13}^{2}$($10^{-3}$eV$^{2}$) (IH) & \quad $2.48$ & \quad $%
2.48_{-0.06}^{+0.05}$ \\ \hline
$\sin ^{2}\theta _{12}$ (IH) & \quad $0.323$ & \quad $0.323\pm 0.016$ \\ 
\hline
$\sin ^{2}\theta _{23}$ (IH) & \quad $0.573$ & \quad $%
0.573_{-0.043}^{+0.025} $ \\ \hline
$\sin ^{2}\theta _{13}$ (IH) & \quad $0.0240$ & \quad $0.0240\pm 0.0019$ \\ 
\hline
\end{tabular}%
\end{center}
\par
\caption{Model and experimental values of the charged lepton masses,
neutrino mass squared splittings and leptonic mixing parameters for the
normal (NH) and inverted (IH) mass hierarchies.}
\label{Observables0}
\end{table}
Using the best-fit values given above, I get for NH and IH, respectively,
the following neutrino masses: 
\begin{eqnarray}
m_{1}&=&0,\hspace{0.5cm}m_{2}\approx 9\mbox{meV},\hspace{0.5cm}m_{3}\approx 50\mbox{meV},\hspace{0.5cm}\mbox{for NH}  \label{neutrinomassesNH}\notag\\
m_{1}&\approx&49\mbox{meV},\hspace{0.5cm}m_{2}\approx 50\mbox{meV},\hspace{0.5cm}m_{3}=0,\hspace{0.5cm}\mbox{for IH}\label{neutrinomassesIH}
\end{eqnarray}%
The obtained and experimental values of the observables in the lepton sector
are shown in Table \ref{Observables0}. The experimental values of the
charged lepton masses, which are given at the $M_{Z}$ scale, have been taken
from Ref. \cite{Bora:2012tx}, whereas the experimental values of the neutrino mass squared splittings and leptonic mixing angles for both normal (NH) and inverted (IH)
mass hierarchies, are taken from Ref. \cite{Forero:2014bxa}. The obtained
charged lepton masses, neutrino mass squared splittings and lepton mixing
angles are in excellent agreement with the experimental data for both normal
and inverted neutrino mass hierarchies. For the sake of simplicity, I assumed all leptonic parameters to be real, but a non-vanishing CP violating phase in the PMNS mixing matrix can be generated by making one of the entries of the neutrino mass matrix $M_{\nu}$ to be complex. 
It is noteworthy to mention that the consistency of the Higgs couplings to SM fermions and gauge bosons with the SM expectation requires that the mixing between the $126$ GeV Higgs and the SM scalar singlet $\chi$ to be  suppressed. In addition, the model cutoff has to be of the order of few TeVs to prevent the one loop level masses and Yukawa couplings for charged fermions lighter than the top quark to be arbitrary small. An ultraviolet completion of this model will consist in replacing the EW scalar singlets $\eta =\left( \eta _{1},\eta _{2}\right) $ by two $SU(2)$ inert scalar doublets unified in a $S_3$ doublet. This will make the model cutoff very high thus avoiding large effects in the low energy flavour physics (FCNC, etc).
%

\section{Conclusions}

I have proposed the first multiscalar singlet extension of the SM that
generates one and three loop level masses for charged fermions lighter than
the top quark and for the light active neutrinos, respectively, via
radiative seesaw mechanisms, without invoking electrically charged scalar
fields. 
These one and three loop radiative seesaw mechanisms are mediated by the $%
\eta _{1}$ and $\eta _{2}$ scalar fields as well as by exotic charged
fermions and right handed Majorana neutrinos, respectively. The model has the $S_{3}\times Z_{8}$ discrete symmetry, where $S_{3}$ is
preserved and $Z_{8}$ is spontaneously broken. 
The breaking of the $Z_{8}$ discrete group generates the non SM fermion
masses as well as the observed SM charged fermion mass and quark mixing
pattern. 
The unbroken $S_{3}$ symmetry of the model allows for natural dark matter candidates, which 
could be either the right handed Majorana neutrinos $\nu _{sR}$ ($s=1,2$) or
the lightest of the $S_{3}$ charged scalar fields $Re(\eta _{1})$, $Re(\eta
_{2})$, $Im(\eta _{1})$ and $Im(\eta _{2})$. 
This possibility is left beyond the scope of this Letter.\vspace{-0.3cm} 
\begin{acknowledgements}
I am very grateful to Professor Riccardo Barbieri for the careful reading of the Letter.
\end{acknowledgements}\vspace{-0.5cm} 

\end{document}